\documentclass[fleqn,twoside,nofootinbib,unicode, twocolumn, hyperref=dvip]{revtex4}
\usepackage[dvips]{graphicx,psfrag}
\usepackage{amssymb}
\usepackage{amsmath}
\usepackage{latexsym}
\usepackage{graphicx}
\usepackage{epsfig}

\begin{document}
\title{Holographic Dynamics as Way to Solve the Basic Cosmological Problems}
\author{Yu.\,L.\,Bolotin}
\email{ybolotin@gmail.com} \author{O.\,A.\,Lemets}\email{oleg.lemets@gmail.com}
\author{D.\,A.\,Yerokhin}\email{denyerokhin@gmail.com}
\author{L.G. Zazunov}
\email{zazun@online.kharkiv.com}
 \affiliation{A.I.Akhiezer Institute for
Theoretical Physics, National Science Center "Kharkov Institute of
Physics and Technology", Akademicheskaya Str. 1, 61108 Kharkov,
Ukraine}

\begin{abstract}
We review recent results on the cosmological models based on the holographic principle which were proposed to explain the most of the problems occurring in the Standard Cosmological Model. It is shown that these models naturally solve the cosmological constant problem and coincidence problem. Well-documented cosmic acceleration at the present time was analyzed in the light of holographic dark energy. In particular, we showed that in the model of Universe consisting of dark matter interacting with a scalar field on the agegraphic background can explain the transient acceleration. We also study the impact of ideas on the physics of entangled states in these cosmological models. Entanglement entropy of the universe gives holographic dark energy with the equation of state consistent with current observational data.
\end{abstract}
\maketitle

\section{Introduction}
One of the most promising ideas that emerged in theoretical physics during the last decade
was the Holographic Principle proposed by 't Hooft and Susskind \cite{tHooft,Susskind,LPSTU,Witten,Maldacena_Gravity}; it appears to be a new guiding paradigm for the true understanding of quantum gravity theories.
Basically, the Holographic Principle states that the fundamental degrees of freedom of a
physical system are bound by its surface area in Planck units.

It concerns the number of degrees of freedom in nature and states that the entropy of matter systems is drastically reduced compared to conventional quantum field theory. This claim is supported by the covariant entropy bound \cite{Bousso1} which is valid in a rather general class of spacetime geometries. The notion of holography is well developed in certain models and backgrounds, in particular in the
context of the $AdS/CFT$ correspondence. A more general formulation is
lacking, however, and the ultimate role of the holographic principle in
fundamental physics remains to be identified.

Although this holographic nature of our description of the particles appears to apply
only for particles entering a black hole, one may argue that it must have a much more
universal validity. According to general relativity, there should exist a direct mapping
that relates physical phenomena in one setting (with a gravitational field present) to
another one (freely falling coordinates)\cite{tHooft}.

 The holographic principle is composed of the two main statements:
\begin{enumerate}
\item The number of possible states of a region of space is the same as that of a system of binary degree of freedom distributed on the boundary of the region, i.e. physics inside a causal horizon can be described completely
by physics on the horizon;
\item The number of such degrees of freedom $N$ is not indefinitely large but is bounded by the area $A$ of the region  (on causal horizon) in Planck units:
\begin{equation}\label{N}
N\le \frac{A{{c}^{3}}}{G\hbar }.
\end{equation}
\end{enumerate}
Therefore holography says that in a quantum theory of gravity we should be able to describe physics in some region
of space by a theory with at most one degree of freedom per unit
Planck area. Notice that the number of degrees of freedom $N$
would then increase  with the area and not with the volume as we are normally
used to. Of course, for all physical systems that we normally  encounter
 the number of degrees of freedom is much smaller than the
area, since the Planck length is so small \cite{Maldacena_Gravity}.
It is called ``holography'' because it would be analogous to a hologram
which can store a three dimensional image in a two dimensional surface.

\section{Cosmological constant and holographic principle }

Why cosmological constant observed today is so much smaller than the
Planck scale? This is one of the most important problems in modern
physics. In history, Einstein first introduced the cosmological
constant in his famous field equation to achieve a static universe
in 1917.
\subsection{The basic problems of the Standard Cosmological Model}
Recent observations of supernov\ae\ \cite{p99}, CMB anisotropies \cite{cmb} and
large scale structure \cite{lss} point to the presence of a flat
universe with a dark energy component.
Understanding the origin of dark energy is one of the
most important challenges facing cosmology and theoretical
physics. One aspect of the problem is
to understand what is the role of zero-point vacuum
fluctuations.

In particle physics, the cosmological constant naturally arises as
an energy density of the vacuum, which is evaluated by the sum of
zero-point energies of quantum fields with mass $m$  as follows
\begin{equation}\label{rho_Lambda}
\rho_\Lambda={1\over2}\int_0^\Lambda{4\pi k^2 dk\over
(2\pi)^3}\sqrt{k^2+m^2}\approx{\Lambda^4\over16\pi^2} ,
\end{equation}

where $\Lambda\gg m$ is the UV cutoff. Usually the quantum field theory
is considered to be valid just below the Planck scale: $M_p\sim
10^{18}$GeV, where we used deduced Planck mass $M_p^{-2}=8\pi G$ for
convenience. If we pick up $\Lambda=M_p$, we find that the energy
density of the vacuum in this case is estimated as $10^{70}$GeV$^4$,
which is about $10^{120}$ orders of magnitude larger than the
observation value $10^{-47}$GeV$^4$. This problem is called the cosmological constant
problem~\cite{lambdarefs}.

Another related but distinct difficulty with
$\Lambda$CDM is the so-called ``why now?'' or coincidence problem.
 \emph{Why} the densities of dark energy and dark matter are comparable \emph{today}?
 While a cosmological constant is by definition time-independent, the matter energy density is diluted as $1/a^{3}$ as the Universe expands. Thus, despite evolution of $a$ over many orders of magnitude, we appear to live in an era during which the two energy densities are roughly the same.
In other words, if $\Lambda$ is tuned to give $\Omega_{\Lambda}\sim\Omega_M$ today, then for
essentially all of the previous history of the universe, the cosmological
constant was negligible in the dynamics of the Hubble expansion, and for the
indefinite future, the universe will undergo a de Sitter-type expansion in
which $\Omega_{\Lambda}$ is near unity and all other components are negligible.
 The present epoch would then be a very special time in the history of the universe,
the only period when $\Omega_M\sim\Omega_{\Lambda}$.

\begin{figure}\label{CCP}
\includegraphics[width=0.45\textwidth]{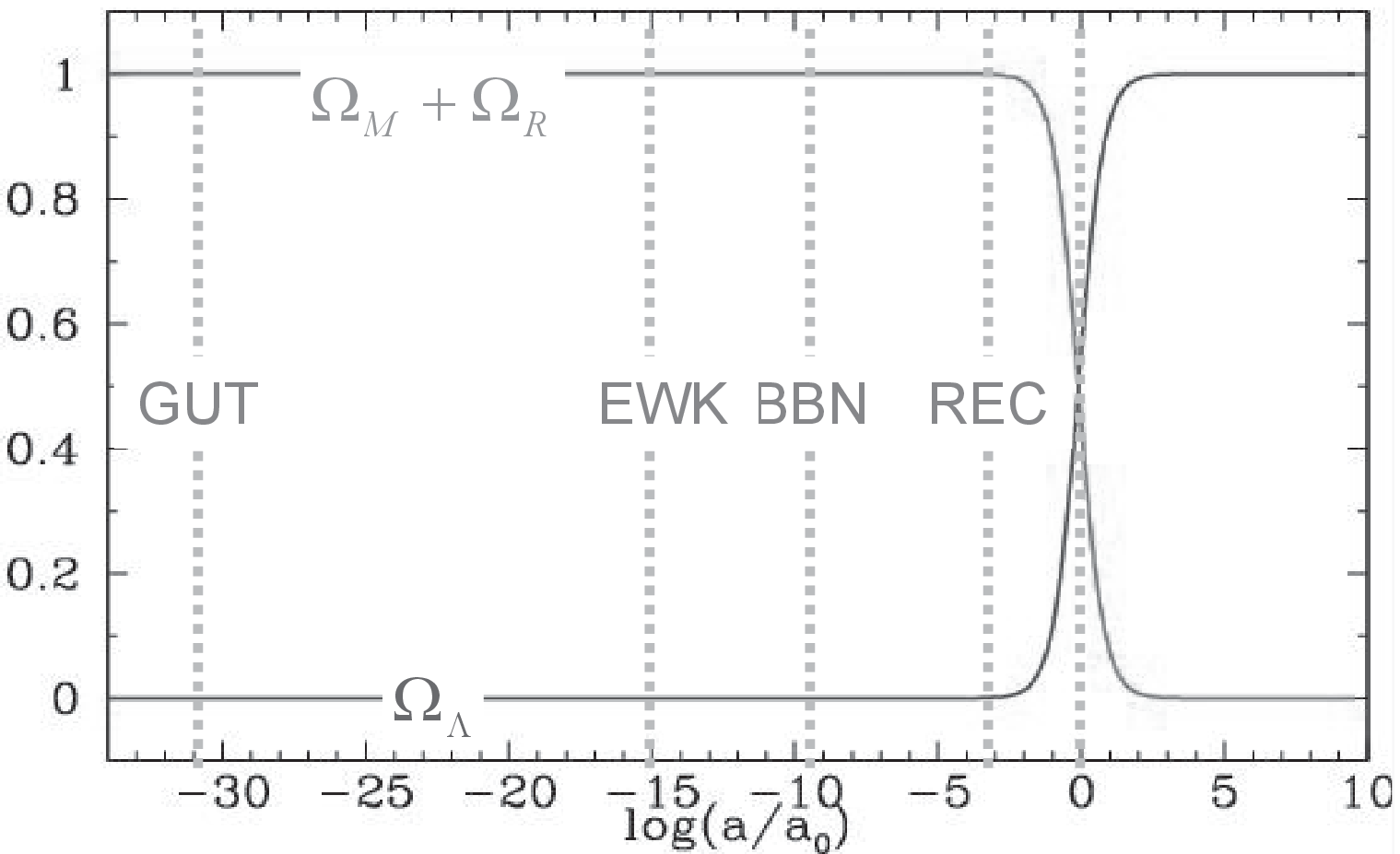}
\caption{The 
change of the vacuum energy density parameter,
  $\Omega_\Lambda,$ as a function of the scale factor $a,$ in
  a universe with $\Omega_{\Lambda 0} = 0.7$, $\Omega_{m0} = 0.3$.
  Scale factors corresponding to the Planck era, electroweak symmetry
  breaking (EW), and Big Bang nucleosynthesis (BBN) are indicated, as
  well as the present day.
  The spike reflects the fact that, in such a universe, there is only
  a short period in which $\Omega_\Lambda$ is evolving noticeably
  with time.}
\end{figure}

\subsection{Cosmological constant:  holographic view}
We now turn to the question of whether some form of a holographic
bound may apply to a cosmological theory in which no boundary conditions have been enforced.

For an effective quantum field theory in a box of size $L$ with UV
cutoff $\Lambda$ the entropy $S$ scales extensively, $S \sim L^3\Lambda^3$

However the specific thermodynamics of black
holes~\cite{Bekenstein,Hawking} has led Bekenstein~\cite{Bekenstein}
to postulate that the maximum entropy in a box of volume $L^3$ behaves non-extensively, growing only as the
area of the box. For any $\Lambda$, there is a sufficiently large
volume for which the entropy of an effective field theory  will exceed
the Bekenstein limit.

The Bekenstein entropy bound may be satisfied in an effective field theory
if we limit the volume of the system according to
\begin{equation}
  \label{bekbound}
  L^3 \Lambda^3 \lesssim S_{BH} \equiv \pi L^2 M_P^2
\end{equation}
where $S_{BH}$ is the entropy of a black hole of radius
$L$~\cite{Hawking,Bekenstein}.
Consequently the length $L$, which acts as an IR cutoff, cannot be chosen
independently of the UV cutoff, and scales as $\Lambda^{-3}$.

An effective field theory that can saturate \eqref{bekbound}
necessarily includes many states with Schwarzschild radius much larger than
the box size.

To avoid these difficulties in \cite{Cohen}  has been proposed
an even  stronger constraint on the IR cutoff $1/L$ which excludes all states that lie within their Schwarzschild radius.  Since the maximum energy density in the effective theory is
$\Lambda^4$, the constraint on $L$ is
\begin{equation}
\label{blueberunium}
L^3 \Lambda^4 \lesssim L M_P^2.
\end{equation}
Here the IR cutoff scales like $\Lambda^{-2}$. This bound is far more
restrictive than \eqref{bekbound}.

While saturating this inequality by choosing the largest $L$ it gives rise to
a holographic energy density
\begin{equation}
\label{rho_de}
\rho_{\Lambda}=3c^{2}M_{p}^{2}L^{-2},
\end{equation}
where $c$ is a dimensionless constant. Then the key issue is
what possible physical scale one can choose as the cutoff $L$
 constrained by the fact of the current acceleration of the universe.

 Applying \eqref{rho_de} this  relation to the Universe as whole it is naturally to identify
the IR-scale with the Hubble radius (simplest case) $H^{-1}.$ Then for the upper
bound of the energy density one finds
 \begin{equation}\label{LsimLambda{-2}}
 \rho_\Lambda\sim L^{-2}M_{Pl}^{2}\sim H^2M_{Pl}^{2}.
 \end{equation}
We will below denote its density as $\rho_{_{DE}}.$ Accounting
that $ M_{Pl} \simeq  1.2\times {{10}^{19}} GeV; ~~~  H_0 \simeq  1.6\times 10^{-42} GeV, $
one finds
\begin{equation}\label{rho_obth}
\rho_{_{DE}}\simeq 3\times 10^{-47}GeV^4 .
\end{equation}
So, this value is comparable to the observed dark energy density $\rho_{obs}\sim 10^{-46}\,GeV^4.$
Therefore the holographic dynamics is free from the cosmological constant
problem.

The coincidences problem  can also be solved within the framework of holographic cosmology \cite{pavon}.
Setting $L =H^{-1}$ in the equation  \eqref{rho_de} and working with the equality (i.e.,
assuming that the holographic bound is saturated) it becomes
$\rho_{_{DE}} = 3\, c^{2}M^{2}_{P}H^{2}$.
Let us consider the flat universe consisting of nonrelativistic matter and holographic dark energy.
The Friedmann equation in this case take the following form $$3M^{2}_{P}H^{2} = \rho_{_{DE}} + \rho_{m}$$ and can be re-written in a very neat form
\begin{equation}\label{concid_holo}
\rho_{m} = 3\left(1 - c^{2}\right)M^{2}_{P}H^{2}.
\end{equation}

 Now, the argument runs as follows: The energy density $\rho_{m}$ varies as
$H^2$, which coincides with the dependence of $\rho_{_{DE}}$ on $H.$
The energy density of cold matter is known to scale as $\rho_{m}
\propto a^{-3}$. So, theirs ratio is constant and has the form
\begin{equation}\label{r}
 \frac{\rho_{m}}{\rho_{DE}} = \frac{1 - c^2}{c^2}.
 \end{equation}
Therefore the holographic dynamics is free from the cosmic coincidences problem
problem.
If ${{\rho }_{_{DE}}}\propto {{H}^{2}},$ then dynamical behavior of holographic dark energy
coincides with that of normal matter, thus the accelerated expansion
is impossible.

In order to produce the accelerated expansion of Universe in frames
of holographic dark energy model we will try to use the IR-cutoff
spatial scale different from the Hubble radius.
The first thing that comes to mind is a consideration as the cutoff value of
the cosmological particle or event horizon.

The particle horizon is given by
\begin{equation}\label{R_H}
R_h=a\int_0^t{dt\over a}=a\int_0^a{da\over Ha^2}.
\end{equation}

Substituting in \eqref{rho_de}  expression for $R_H$ and using the equation of covariant conservation,  it is easy to verify \cite{MiaoLi} that  expression for the equation of state parameter $w=p/\rho$ takes the form $w=-{1\over 3}+{2\over 3c}>-{1\over3}.$ So, this IR-scale can not provide  the accelerated expansion of the universe. To get an accelerating
expansion of Universe, we supersede the particle horizon by the future event horizon
$$R_h=a\int_t^\infty{dt\over a}=a\int_a^\infty {da\over
Ha^2}.$$ This horizon is the boundary of the volume a fixed
observer may eventually observe. One is to formulate a theory
regarding a fixed observer within this horizon.
In this case, the equation of state parameter acquire the form
\begin{equation}\label{IR_EH}
w=-{1\over 3}-{2\over 3c}.
\end{equation}

We  obtain a component of energy behaving as dark energy. If we take $c=1$,
its behavior is similar to the cosmological constant. If $c<1$,
$w<-1$, a value achieved in the past only in the phantom model in the traditional approach.

For the first impression the declared task is completed. The
holographic dark energy with equation of state parameter \eqref{IR_EH} from the one
hand provides correspondence between the observed density and the
theoretical estimate, and from the other it leads to the state
equation which is able to generate the accelerated expansion of
Universe. However the holographic dark energy with IR-cutoff on the
event horizon still leaves unsolved problems connected with the
causality principle: according to the definition of the event
horizon the holographic dark energy dynamics depends on future
evolution of the scale factor. Such dependence is hard to agree with
the causality principle.

In addition to these parameters of infrared cutoff scale in the literature are discussed the Ricci scalar curvature and scale associated with the age of the universe. The latter type of  dark energy is so-called agegraphic dark energy.

The former based on the fact that space-time curvature is non-zero, so it can be associated with a horizon, that is considered as a holographic screen. The latter type of   energy is so-called the agegraphic dark energy. This kind of dark energy we study in more detail.
\subsubsection{Agegraphic dark energy}
According to the definition the agegraphic dark energy is the holographic dark energy in the infrared cutoff scale equal to the age of the universe. It is remarkable that this kind of energy can be obtained from independent and less radical conception.

The existence of quantum fluctuations in the metric
 directly leads to the following conclusion, related to the problem of distance
measurements in the Minkowski space: the distance $t$
\footnote{recall that we use the system of units where the light
speed equals $c=\hbar =1,$ so that
${{L}_{Pl}}={{t}_{Pl}}=M_{Pl}^{-1}$} cannot be measured with
precision exceeding the following
\begin{equation}
\label{Karolyhazy_rel} \delta t=\beta t_{Pl}^{2/3}{{t}^{1/3}},
\end{equation}
where $\beta $ is a factor of order of unity.
This expression is  so-called Karolyhazy uncertainty relation \cite{Karolyhazy}.

The K\'arolyh\'azy relation (\ref{Karolyhazy_rel}) together with the
time-energy uncertainty relation enables one to estimate a quantum
energy density of the metric fluctuations of Minkowski
space-time~\cite{Maziashvili_1,Maziashvili_2,0707.4049}. With the relation (\ref{Karolyhazy_rel}), a length scale $t$ can be known with a maximal precision $\delta t$
determining a minimal detectable cell $\delta^3$ over a spatial
region $t^3$. Thus one is able to look at the microstructure of
space-time over a region $t^3$ by viewing the region as the one
consisting of cells $\delta t^3 \sim t_p^2t$. Therefore such a cell
$\delta t^3$ is the minimal detectable unit of space-time over a
given length scale $t$ and if the age of the space-time is $t$, its
existence due to the time-energy uncertainty relation cannot be
justified with energy smaller than $\sim t^{-1}$ . Hence, as a
result of the relation (\ref{Karolyhazy_rel}), one can conclude that if the age
of the Minkowski space-time is $t$ over a spatial region with linear
size $t$ (determining the maximal observable patch) there exists a
minimal cell $\delta t^3$, the energy of the cell cannot be smaller
than~\cite{Sasakura,Maziashvili_2,0707.4049}
 \begin{equation}
  \label{eq02}
  E_{\delta t^3} \sim t^{-1},
  \end{equation}
  due to the time-energy uncertainty relation, it was argued  that the energy density of metric fluctuations of  Minkowski spacetime is given by
 \begin{equation}
\label{rho_q}
 \rho_q\sim\frac{E_{\delta t^3}}{\delta t^3}\sim
 \frac{1}{t_p^2 t^2}\sim\frac{M_p^2}{t^2}.
\end{equation}

In~\cite{Maziashvili_2} (see also~\cite{0707.4049}), it is noticed that the
 K\'{a}rolyh\'{a}zy relation naturally obeys the  holographic black hole entropy bound \cite{Bekenstein}.  It is worth noting that  the form of energy density Eq.~(\ref{rho_q}) is similar to the one  of holographic dark energy~\cite{0707.2129,0701405}, i.e.,  $\rho_\Lambda\sim l_p^{-2}l^{-2}$. The similarity between  $\rho_q$ and $\rho_\Lambda$ might reveal some universal features  of quantum gravity, although they arise from different sources.
As the most natural  choice, the time scale $t$ in Eq.~(\ref{rho_q}) is chosen to
 be the age of our Universe. Therefore, we call it ``agegraphic'' dark  energy \cite{0707.4049}.

The relation \eqref{rho_q} allows to introduce an alternative model
for holographic dark energy, which uses the age of
Universe $T$ for IR-cutoff scale. In such a model
\begin{equation}\label{rho_qn}
\rho_q=\frac{3{{n}^{2}}M_{Pl}^{2}}{{{T}^{2}}},
\end{equation}
where $n$ is a free parameter of model, and the number coefficient
$3$ is introduced for convenience. So defined energy density
\eqref{rho_qn} with $T\sim H_{0}^{-1},$ where ${{H}_{0}}$ is the
current value of the Hubble parameter, leads to the observed value
of the dark energy density with the coefficient $n$ value of order
of unity. Thus in SCM, where ${{H}_{0}}\simeq 72km\,{{\sec
}^{-1}}Mpc^{-1},\ {{\Omega }_{DE}}\simeq 0.73,\ T\simeq
13.7\,Gyr,$ one finds that $n\simeq 1.15.$

Suppose that the Universe is described by the Friedmann equation
\begin{equation}\label{H2rho_qnpho_m}
{{H}^{2}}=\frac{1}{3M_{Pl}^{2}}\left( {{\rho }_{q}}+{{\rho }_{m}}
\right).
\end{equation}
The state equation for the dark energy is
\begin{equation}\label{w_qOmega_q}
{{w}_{q}}=-1+\frac{2}{3n}\sqrt{{{\Omega}_{q}}}.
\end{equation}
So such  universe will be accelerated expanded, and would be similar to $\Lambda CDM.$
Thus the holographic model for dark energy with IR-cutoff
scale set to the Universe age, allows the following:
\begin{enumerate}
  \item to obtain the observed value of the dark energy density;
  \item provide the accelerated expansion regime on later stages of
  the Universe evolution;
  \item resolve contradictions with the causality principle.
\end{enumerate}
The first successes of holographic principle application from one hand awoke hopes to create on that basis an adequate description of the Universe dynamics, free of a number of problems which the traditional approach suffers from.
Nevertheless we do not discard other possibilities, on which exist  the  allusions of observations.

\subsection{Observations Challenge}
\begin{figure}[t]
\centering\label{fig_Srarob}
\includegraphics[width=0.47\textwidth]{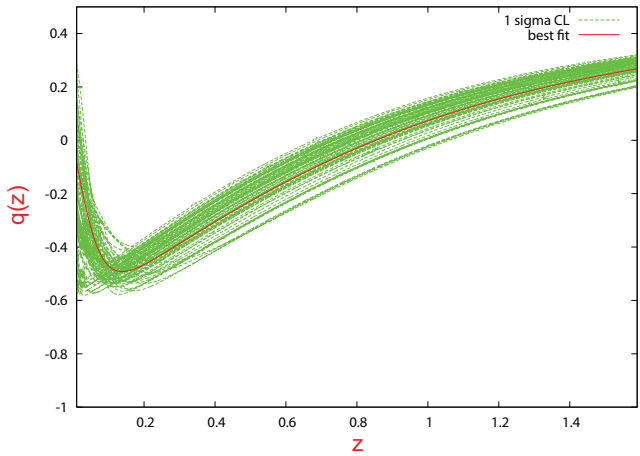}
\caption{The deceleration parameter dependence $q(z)$ reconstructed
from independent observational data, including the brightness curves
for SN Ia, cosmic microwave background temperature anisotropy and
baryon acoustic oscillations (BAO). The red solid line shows the best fit on the
confidence level $1\sigma$ CL\cite{Starobinsky}.}
\end{figure}

A. Starobinsky \cite{Starobinsky} and co-workers investigated the course of cosmic expansion in its  recent past using the Constitution SN Ia sample (which includes CfA data at low redshifts), jointly with
signatures of baryon acoustic oscillations (BAO) in the galaxy distribution
 and fluctuations in the cosmic microwave background (CMB). Allowing the equation of state of dark energy (DE) to vary, they find that a coasting model of
the universe ($q_0=0$) fits the data about as well as $\Lambda$CDM. This effect, which is most clearly seen using the recently introduced $Om$ diagnostic \cite{Starobinsky_Om}, corresponds to an increase of
$Om$ and $q$ at redshifts $z\lesssim 0.3$. In geometrical terms, this suggests
that cosmic acceleration may have already peaked and that we are currently witnessing its slowing down (see figure \ref{fig_Srarob}).

The case for evolving dark energy strengthens if a subsample of the Constitution set consisting of SNLS+ESSENCE+CfA SN Ia data is analysed in combination with BAO+CMB using the same statistical methods.
Thus the main result of the analysis is the following: SCM is not unique though the simplest explanation
of the observational data, and the accelerated expansion of Universe
presently dominated by dark energy is just a transient phenomenon.

Thus, the evolutional behavior of dark energy reconstructed and the issue of whether
the cosmic acceleration is slowing down or even speeding up is highly dependent upon the SNIa data sets, the light curve fitting method of the SNIa, and the parametrization forms of the equation of state. In order to have a definite answer, we must wait for data with more precision and search for the more reliable and efficient methods to analyze these data.

Model with the holographic dark energy, as discussed above, in their original form, do not allow to explain the nonmonotonic dependence of the cosmological parameters.

\section{The model of interacting dark energy with a transient acceleration phase}
Current literature usually considers the models where the required
dynamics of Universe is provided by one or another, and always only
one, type of dark energy.

As was multiply mentioned above, in order
to explain the observed dynamics of Universe, the action for
gravitational field is commonly complemented, besides the
conventional matter fields (both matter and baryon), by either the
 cosmological constant, which plays role of physical vacuum in SCM,
 or more complicated dynamical objects --- scalar fields,
$K$-essence and so on. In the context of holographic cosmology, the
latter term is usually neglected, restricting to contribution of the
boundary terms. Nevertheless such restriction has no theoretical
motivation.

We consider the cosmological model which contains both volume and surface terms. The role of former is played
by homogeneous scalar field in exponential potential, which
interact with dark matter. The boundary term responds to
holographic dark energy in form of \eqref{rho_qn}.
 This  scenario   predicts  transient accelerating phase.

To describe the dynamical properties of the Universe it is
convenient to to transform to dimensionless variables as the
following:
\begin{equation}\label{var2}
\begin{array}{c}
x=\frac{\dot{\varphi}}{\sqrt{6}M_{Pl}H},
y=\frac{1}{M_{Pl}H}\sqrt{\frac{V(\varphi)}{3}}, \\
 z=\frac{1}{M_{Pl}H}\sqrt{\frac{\rho_{m}}{3}},
u=\frac{1}{M_{Pl}H}\sqrt{\frac{\rho_{_{q}}}{3}}.
\end{array}
\end{equation}
The evolution of scalar field is described by the Klein-Gordon
equation, which in the case of interaction between the scalar field
and matter takes the following form:
\begin{equation}
\ddot{\varphi}+3 H \dot{\varphi} + \frac{dV}{d\varphi}=-
\frac{Q}{\dot{\varphi}}. \label{eq:kgeqn}
\end{equation}
In the present section we consider the case when the interaction
parameter $Q$ is a linear combination of energy density for scalar
field and dark energy
\begin{equation}
  \label{Q1}
  Q= 3H(\alpha\rho_\varphi+\beta \rho_m),
\end{equation}
where $\alpha,$ $\beta$ are constant parameter. For given model,
regardless the explicit form of the scalar field potential
$V(\varphi).$

As was mentioned above, here we consider the simplest case of
exponential potential
\begin{equation}\label{V_exp}
    V=V_0\exp\left(\sqrt{\frac{2}{3}}\frac{\mu\varphi}{M_{Pl}}\right),
\end{equation}
where $\mu$ is constant.

Taking into account the expression
\eqref{var2}, the system of equations describing the dynamics of the universe in this model reads
\begin{eqnarray}
x' &=& \frac{3x}{2}\left[g(x,z,u)- \frac{\alpha(x^2+y^2)+\beta z^2}{x^2}\right]-3x - \mu y^2,\nonumber\\
y' &=&  \frac{3y}{2}g(x,z,u)+\mu xy, \\ \label{sys_xyz_V}
z' &=&\frac{3z}{2}\left[g(x,z,u)+\frac{\alpha(x^2+y^2)+\beta z^2}{z^2}\right]-\frac{3}{2}z ,\nonumber \\
u'&=& \frac{3u}{2}g(x,z,u)-\frac{u^2}{n},\nonumber
\end{eqnarray}
where
\begin{equation}
g(x,z,u) = 2x^2+ z^2+ \frac{2}{3n}u^3,~\lambda\equiv
-\frac{1}{V}\frac{dV}{d\varphi} M_{Pl}.
\end{equation}

\begin{figure}[t]
\centering
\includegraphics[width=0.47\textwidth]{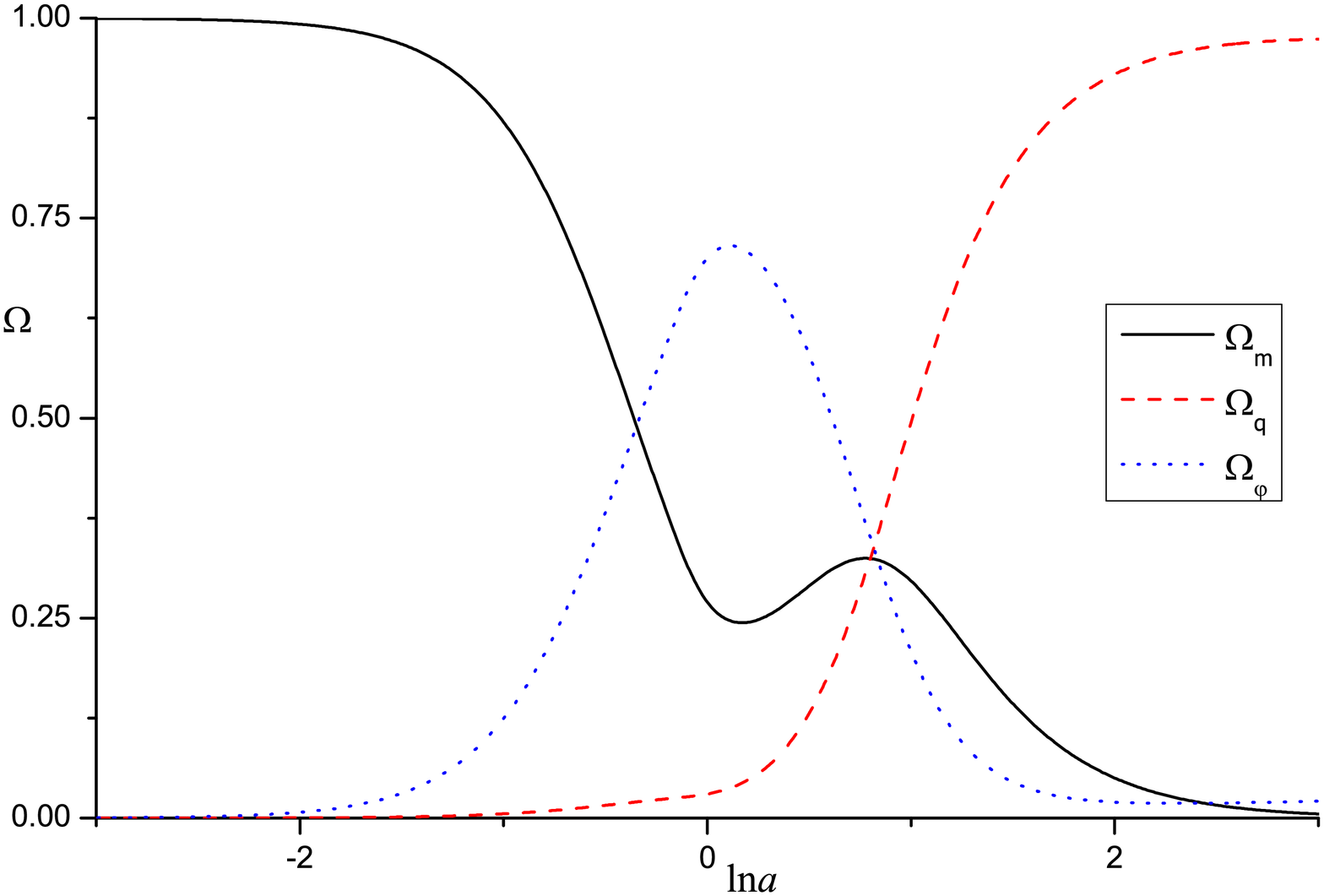}
\includegraphics[width=0.47\textwidth]{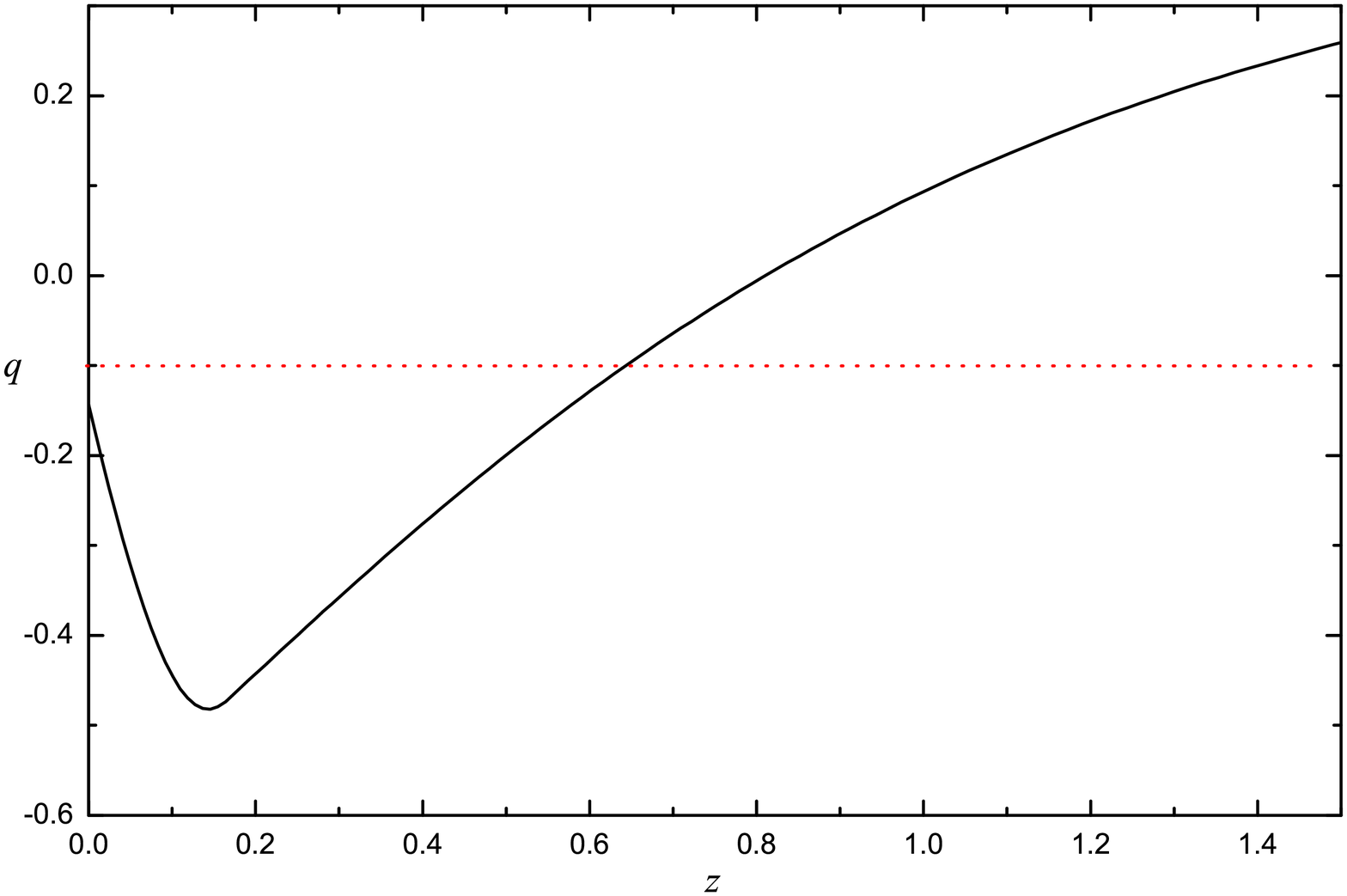}
\caption{Behavior of $\Omega_{\varphi}$ (dot line), $\Omega_{q}$
(dash  line) and $\Omega_{m}$ (solid line) as a function of  $N=\ln
a$ for $n=3,\;\alpha = 0.005$ and $\mu = -5$ (upper plot). Evolution
of deceleration parameter for this model (center) and the deceleration parameter  $q(z)$ reconstructed
from independent observational data (lower plot).}
\label{fig:4}
\end{figure}

Next, we consider the simplest case in which this model can obtained the regime of  transient acceleration.
 $Q=3H\alpha\rho_\varphi$
 We consider the case with the interaction parameter of the form (\ref{Q1}) with $ \beta = 0.$
For example, we consider the case presented in the figure \ref{fig:4}.
With these values of the parameters of interaction, transient acceleration begins almost in the present era.
So, one of the deficiencies of original agegraphic dark energy model is the inability to explain the phenomenon of transient acceleration, in this model can be solved.

\section{Entanglement entropy and holography}
In quantum information science, quantum entanglement
is a central concept
 and a precious resource allowing various
quantum information processing such as quantum key
distribution.
The entanglement  is a  quantum nonlocal
correlation which can not be prepared by  local operations and
classical communication \cite{JaeWeonLee}.
For pure states the entanglement entropy $S_{Ent}$ is a good measure of entanglement.
For a bipartite system $AB$ described by a full density
matrix $\rho_{_{AB}}.$   The von Neumann entropy $S_{Ent}$  is
\begin{equation}\label{S_{Ent}}
S_{Ent}=-Tr\,(\rho_{_A} \ln \rho_{_A}),
\end{equation}
where $\rho_{_A}$ -- reduced matrix obtained by ``tracing out'' the degrees of freedom of system $B$ (which is quantum-correlated with $A$) and given by 
\begin{equation}
\rho_{_A} \equiv Tr_{_B} \rho_{_{AB}}
\end{equation}

\begin{figure}[hbtp]\label{entangl}
\includegraphics[width=0.47\textwidth]{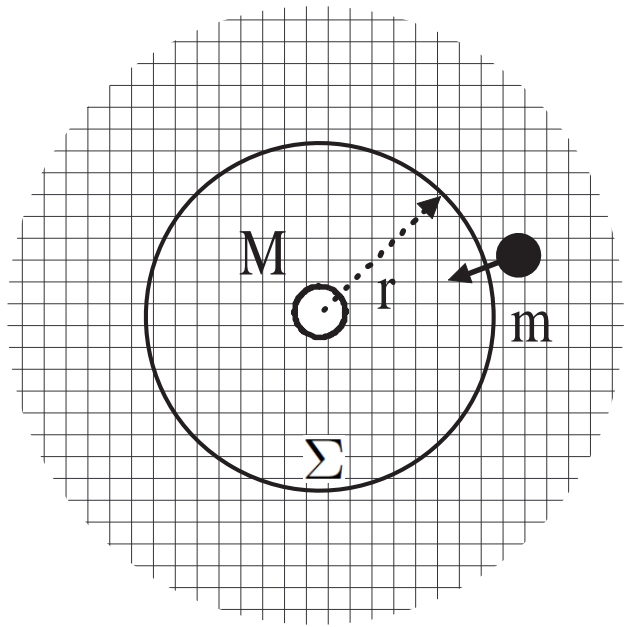}
\caption{ The space around a massive object with mass $M$
can be divided into two subspaces, the inside and the outside of an imaginary spherical
surface with a radius $r$.
The surface $\Sigma$  has the entanglement entropy
$S_{ent}\propto r^2$ and entanglement energy
 $E_{ent}\equiv  \int_\Sigma T_{ent} dS_{ent}$.
If there is a test particle with mass $m$, it feels an effective attractive force
in the direction of  increase of entanglement.
}
\end{figure}
The Basic Conjecture of the Entanglement entropy are
\begin{enumerate}
  \item Quantum entanglement of matter or the vacuum in the universe increases like the entropy;
  \item There is a new kind of force - quantum entanglement force associated with this tendency;
  \item Gravity and dark energy are types of the quantum entanglement force associated with the increase of the entanglement, similar to Verlinde's \cite{Verlinde:2010hp}  entropic force linked with the increase of the entropy.
\end{enumerate}

 For an entanglement system in the flat spacetime, we  consider the
three-dimensional  spherical  volume $V$ and its enclosed boundary $\Sigma$ (See Fig. \ref{entangl}).
 We assume that this system with  radius $r$  and the cutoff scale $b$ is described by  the local quantum
 field theory of a free scalar field $\phi.$

In general, the vacuum entanglement entropy of a spherical region with a radius $r$
 with quantum fields can be expressed in the form
\begin{equation}
 \label{Sent}
  S_{ent}=\frac{\beta r^2}{b^2},
\end{equation}
   where $\beta$ is an $O(1)$ constant that
depends on the nature of the field (like $n$ in the agegraphic dark energy) and $b$ is the UV-cutoff.

The entanglement energy is carried by the modes around ${\cal B}$,
which implies that the cutoff scale $b$ is introduced only in the
$r$ direction through the length contraction $b^2/r^2$. We start by
noting the similarity between the entanglement system in the flat
spacetime and  the stretched horizon formulation of the
Schwarzschild black hole.

So, $S_{Ent}$ has a form consistent with the holographic principle,  although it is derived
from quantum field theory without using the principle.
Thus, from different and independent physical assumptions, we come to equal physical consequence.
We can use both of this ideology with equal success and equivalent effect.

Why are we considering the quantum  entanglement as an essential concept for  gravity?
First, there are interesting similarities between the holographic entropy and the entanglement entropy of a given surface. Both are proportional to its area in general and related to quantum nonlocality.
Second, when there is a gravitational force, there  is always a Rindler horizon for some observers,
which acts as information barrier for the observers.
This can lead to ignorance of information beyond the horizons, and
the lost information can be described by the entanglement entropy.
The spacetime should bend itself so that the increase of the entanglement entropy
compensate the lost information of matter.
Third, if we use the entanglement entropy of quantum fields instead of thermal
entropy of the holographic screen, we can understand the microstates of the
screen and explicitly  calculate, in principle, relevant physical quantities using the quantum field theory
in the curved spacetime. The microstates can
be thought of as  just quantum fields on the surface or its discretized oscillators.
Finally, identifying the holographic entropy as the entanglement entropy could
explain why the derivations of the Einstein equation is involved with entropy, the Planck constant $\hbar$
and, hence, quantum mechanics.
All these facts indicate that quantum mechanics and gravity has an intrinsic connection,
and the holographic principle itself has something to do with quantum entanglement.

\section{Conclusion}
In this work we did a brief overview of the application of holographic principle for solving the basic problems of the standard cosmological model.
It is shown how a model based on the holographic principle naturally solve the cosmological constant problem and coincidence problem. Proposed modification of  this model was capable of explaining the possibility of nonlinear dynamics of the cosmological parameters - the phenomenon transient acceleration. It is shown that there is a deep analogy between the cosmological models with the holographic principle and models with quantum entanglement entropy.

\section*{Acknowledgements}
Work is supported in part by the Joint DFFD-RFBR Grant \# F40.2/040.

\end{document}